\documentstyle[11pt,newpasp,twoside,epsf]{article}
\def\edcomment#1{\iffalse\marginpar{\raggedright\sl#1\/}\else\relax\fi}
\reversemarginpar

\begin{document}

\title {Optical and EUV Light Curves of Dwarf Nova Outbursts}
\author{Christopher W.~Mauche}
\affil {Lawrence Livermore National Laboratory,
        L-43, 7000 East Avenue, Livermore, CA 94550}
\author{Janet A. Mattei}
\affil {American Association of Variable Star Observers,
        25 Birch Street, Cambridge, MA 01238}
\author{Frank M. Bateson}
\affil {Variable Star Section/Royal Astronomical Society of New Zealand,
        P.O.~Box 3093, Greerton, Tauranga, New Zealand}

\begin{abstract}
We combine AAVSO and VSS/RASNZ optical and {\it Extreme Ultraviolet
Explorer\/} EUV light curves of dwarf novae in outburst to place
constraints on the nature of dwarf nova outbursts. From
the observed optical--EUV time delays of $\approx 0.75$--1.5 days,
we show that the propagation velocity of the dwarf nova
instability heating wave is $\approx 3~\rm km~s^{-1}$.
\end{abstract}

\section{Introduction}

Dwarf nova outbursts are understood to be the result of an instability 
in the rate of mass transfer through the accretion disk surrounding the
white dwarf in these semidetached binaries. The instability can be
triggered at large or small disk radii, resulting in normal, fast-rise
outbursts or anomalous, slow-rise outbursts, respectively. In either case,
the beginning of the outburst is signaled by a rise of the optical flux,
followed by a rise of the UV flux as material sinks through the disk,
converting gravitational potential energy into rotational kinetic energy
and radiation. This is followed by a rise in the EUV flux as material
passes through the boundary layer between the disk and the surface of the
white dwarf, where it converts its prodigious rotational kinetic energy
into radiation.

Because the optical and EUV flux of dwarf novae is produced in physically
distinct regions of the accretion disk, light curves in these wavebands
provide important diagnostics of the nature of dwarf nova outbursts.
Toward this end, we present optical and EUV light curves of dwarf novae
observed by the {\it Extreme Ultraviolet Explorer\/} ({\it EUVE\/})
satellite.

\section{EUVE Observations}

During the past seven years, {\it EUVE\/} was used to observe four dwarf
novae in outburst: SS~Cyg in narrow and wide, normal and anomalous
outbursts; U~Gem in normal outburst (twice); VW~Hyi in normal and
super\-outburst (twice); and OY~Car in superoutburst (twice). These eleven
observations were originally obtained for varying reasons, and on a couple
of occasions they were coordinated with other satellites ({\it RXTE\/},
{\it Voyager\/}, {\it HST\/}) sensitive in other wavebands (hard X-rays,
FUV, UV). Details of the {\it EUVE\/} observations are provided in
Table~1. Note that the exposures are 55--274 ks, and because the
satellite takes data for $\la $ one third of its orbit, the EUV light
curves span intervals of 2--13 days.

\begin{table}
\vskip -0.2cm
\caption{Journal of {\it EUVE\/} Observations}
\begin{tabular}{lcccll}
\tableline
    &  Date&	      Interval&	Exp.&	Type       &        \\
Star&	(M/Y)&	(JD$-$2400000)&	(ks)&	of Outburst&	Comment\\
\tableline
SS Cyg&	08/93&	49216.58--222.86&	          179.4&  Anom.~Wide&    \\
 U Gem&	12/93&	49350.00--361.15&	          249.0&  Normal&        \\
VW Hyi&	06/94&	49505.46--507.66&	\phantom{0}89.4&  Super&         \\
SS Cyg&	06/94&	49526.67--536.69&	          147.8&  Normal Wide&   \\
VW Hyi&	07/95&	49906.70--917.29&	          183.8&  Normal&        + {\it Voyager\/}\\
VW Hyi&	05/96&	50210.58--218.47&	\phantom{0}55.4&  Super&         + {\it RXTE\/}   \\
SS Cyg&	10/96&	50366.40--379.45&	          208.1&  Normal Narrow& + {\it RXTE\/}   \\
OY Car&	03/97&	50534.46--537.64&	\phantom{0}94.8&  Super&         \\
 U Gem&	11/97&	50760.27--766.85&	          150.0&  Normal&        + {\it RXTE\/}   \\
SS Cyg&	06/99&	51336.84--349.67&	          274.0&  Anom.~Narrow&  + {\it RXTE\/}\\
OY Car&	02/00&	51597.66--601.26&	\phantom{0}69.1&  Super&         \phantom{+ }{\small \&}{\it HST\/}\\
\tableline
\multicolumn{2}{l}{Net Exposure:}&	87.4 days& 1.7 Ms&      &         \\
\end{tabular}
\vskip -0.2cm
\end{table}

\section{Optical and EUV Light Curves}

Optical light curves of these outbursts were constructed from visual
magnitude estimates and CCD photometric measurements obtained by members
of the American Association of Variable Star Observers (AAVSO) and
the Variable Star Section/Royal Astronomical Society of New Zealand
(VSS/RASNZ). EUV light curves were constructed from the {\it EUVE\/} deep
survey photometer (DS) or short wavelength spectrometer (SW) in those
instances when the DS was turned off (during the peak of the 10/96
outburst of SS~Cyg, both outbursts of U Gem, and the 6/94 outburst of
VW~Hyi). The resulting optical and EUV light curves are shown in Figures
1--6, with the {\it EUVE\/} DS and SW data shown by filled circles and
squares, respectively, and the optical data shown by dots and open 
diamonds (measurements) or carets (upper limits). Half-day averages of the
optical measurements are shown by the histograms for SS~Cyg and U~Gem. In
Figure~4 the {\it Voyager\/} 950--1150~\AA \ (FUV) flux density light
curve of VW~Hyi in normal outburst is shown by the filled triangles.

\begin{figure}
\plottwo{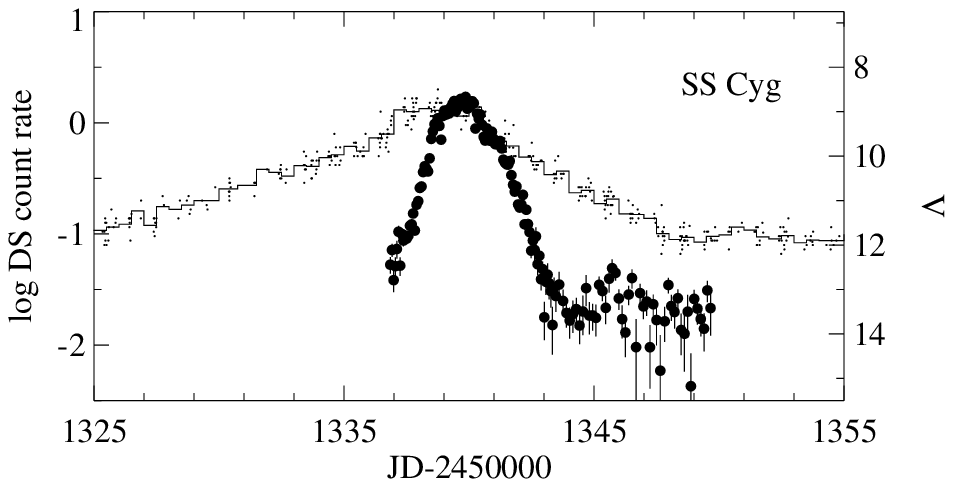}{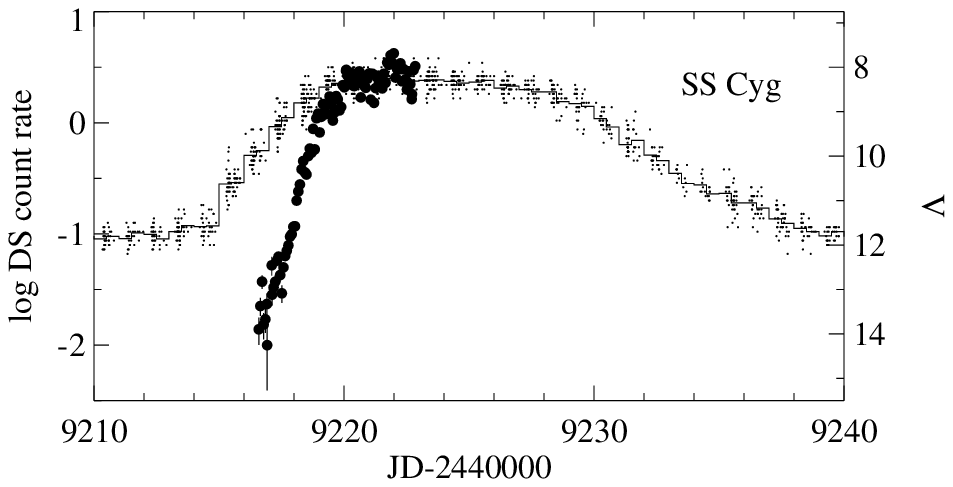}
\caption{Anomalous (slow-rise) outbursts of SS Cyg.}
\end{figure}

\begin{figure}
\plottwo{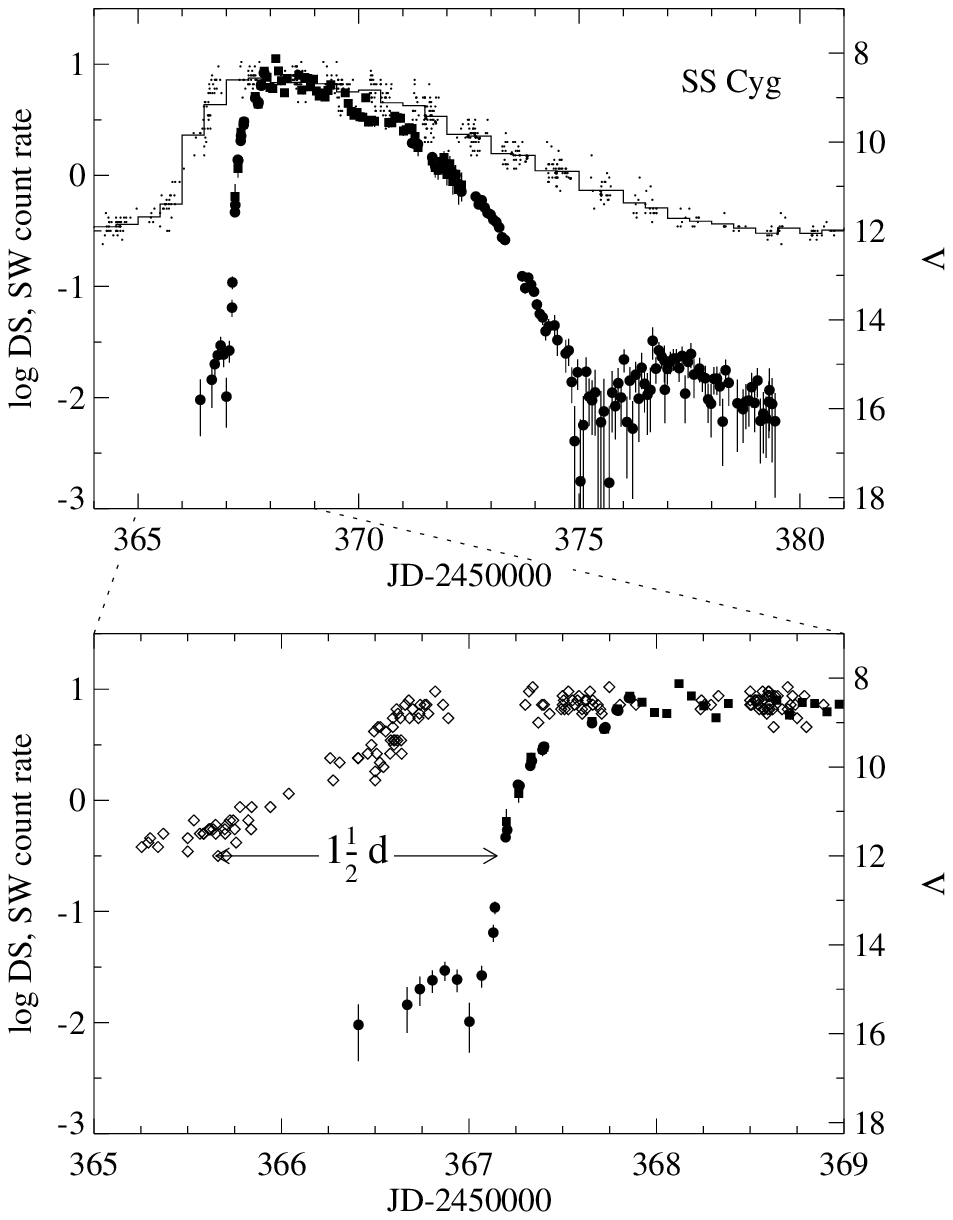}{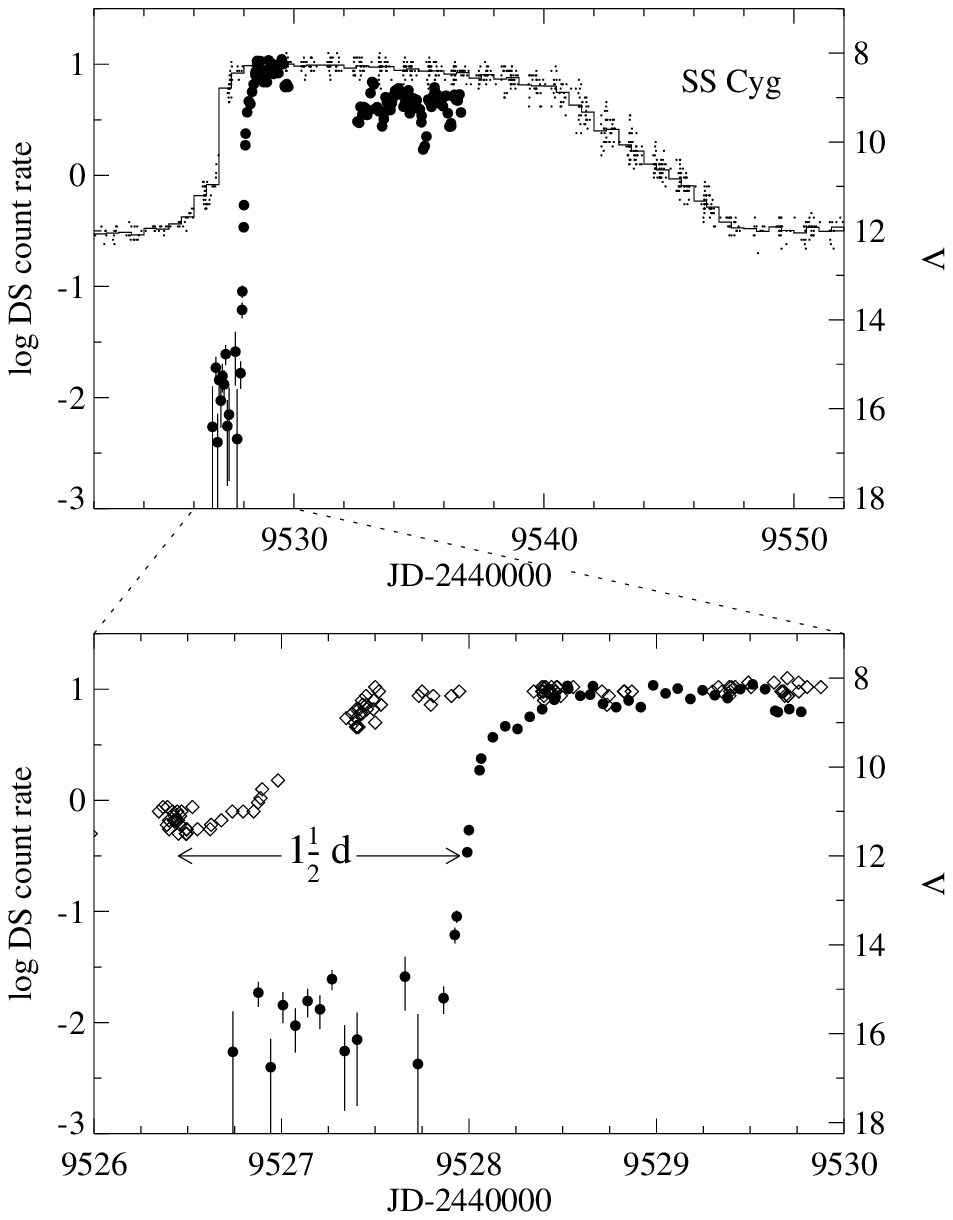}
\caption{Normal (fast-rise) outbursts of SS Cyg. 
The optical-EUV delay is $\approx 1.5$ days.}
\end{figure}

\begin{figure}
\plottwo{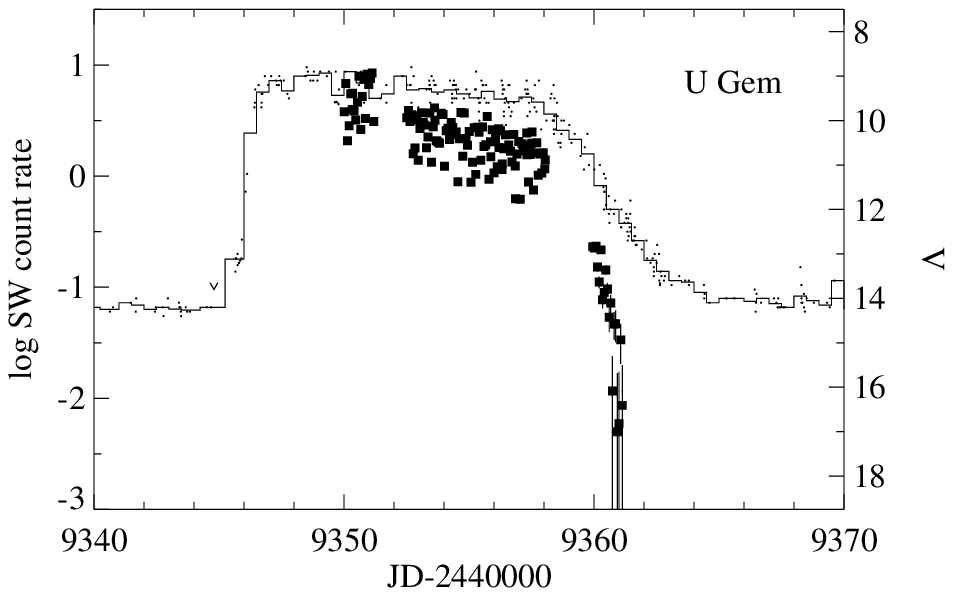}{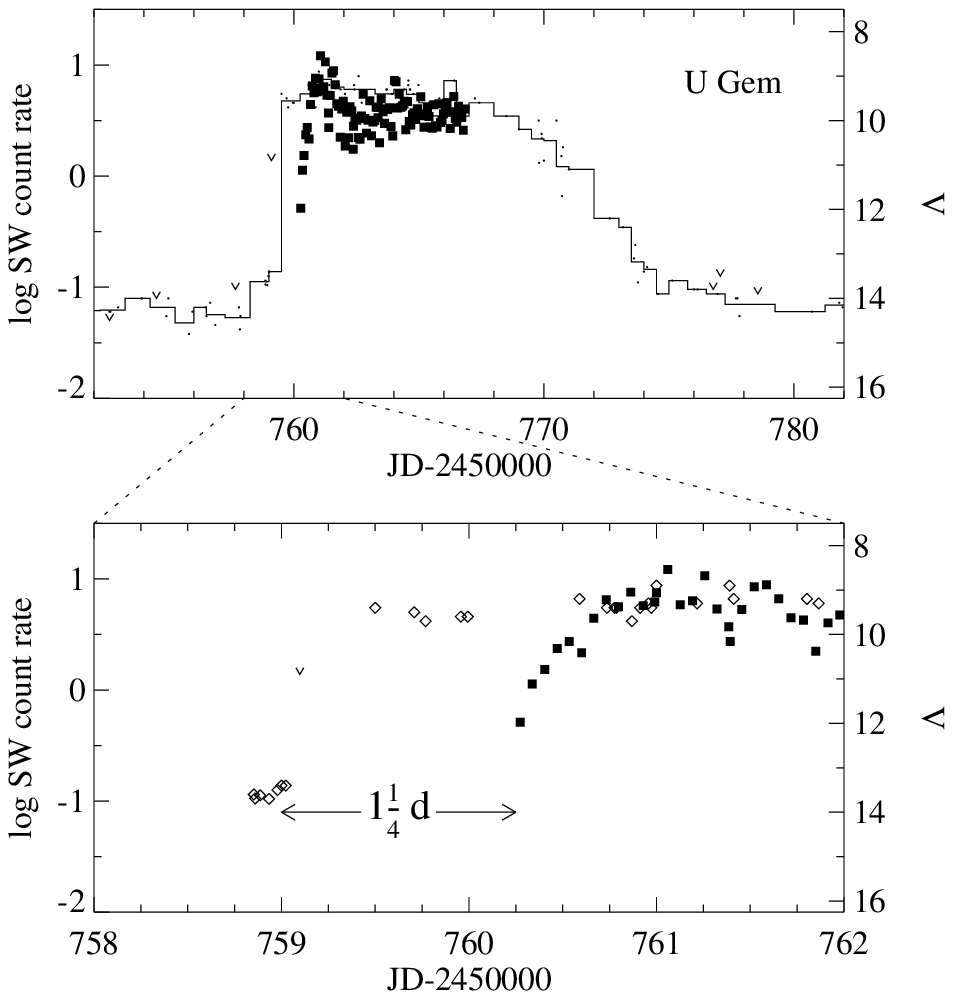}
\caption{Outbursts of U Gem.
The optical-EUV delay is $\approx 1.25$ days.}
\end{figure}

\begin{figure}
\plotfiddle{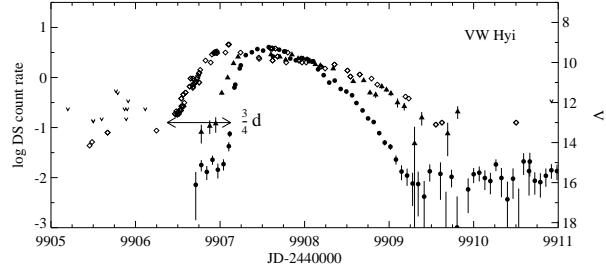}{3.5cm}{0}{59}{59}{-108}{0}
\caption{Normal outburst of VW Hyi.
The optical-FUV delay is $\approx 0.5$ days and the optical-EUV delay
is $\approx 0.75$ days.}
\end{figure}

\begin{figure}
\plottwo{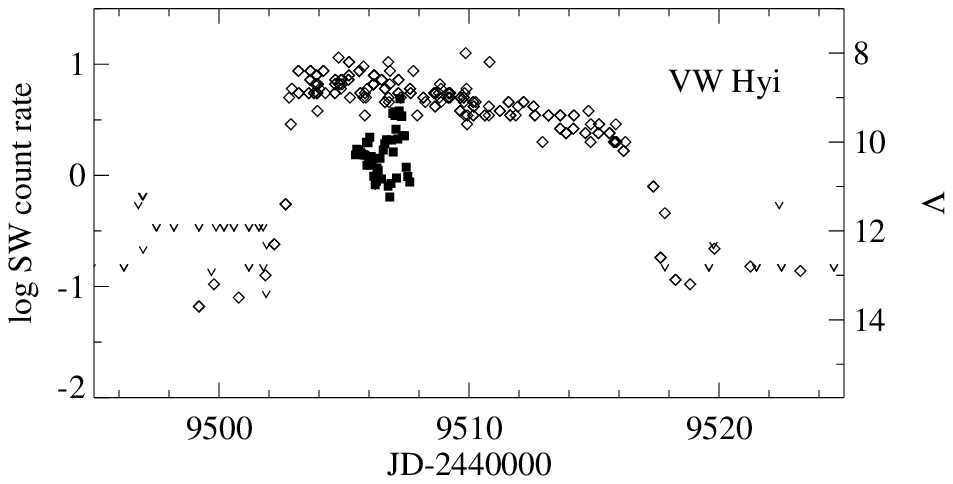}{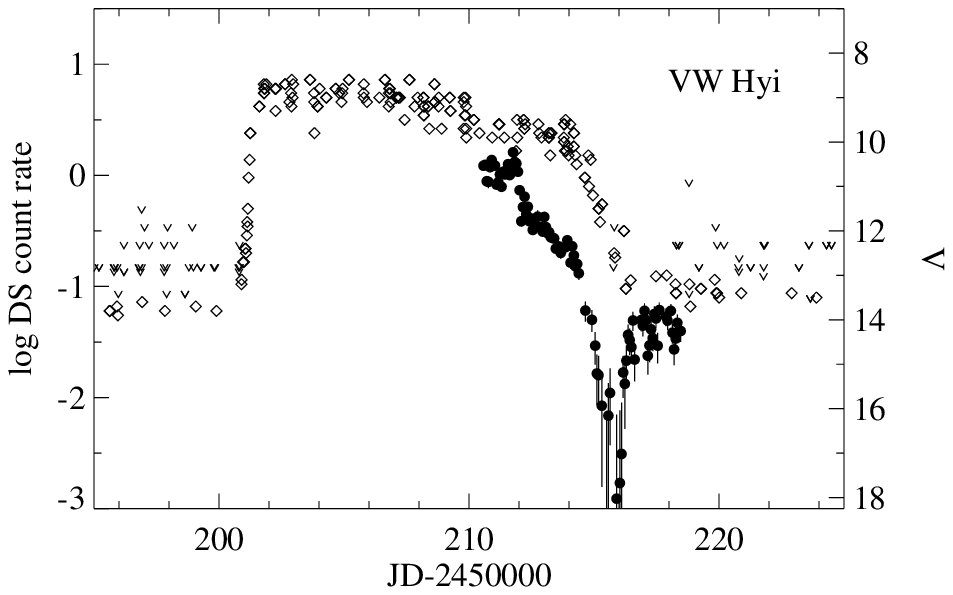}
\caption{Superoutbursts of VW Hyi.}
\end{figure}

\begin{figure}
\plottwo{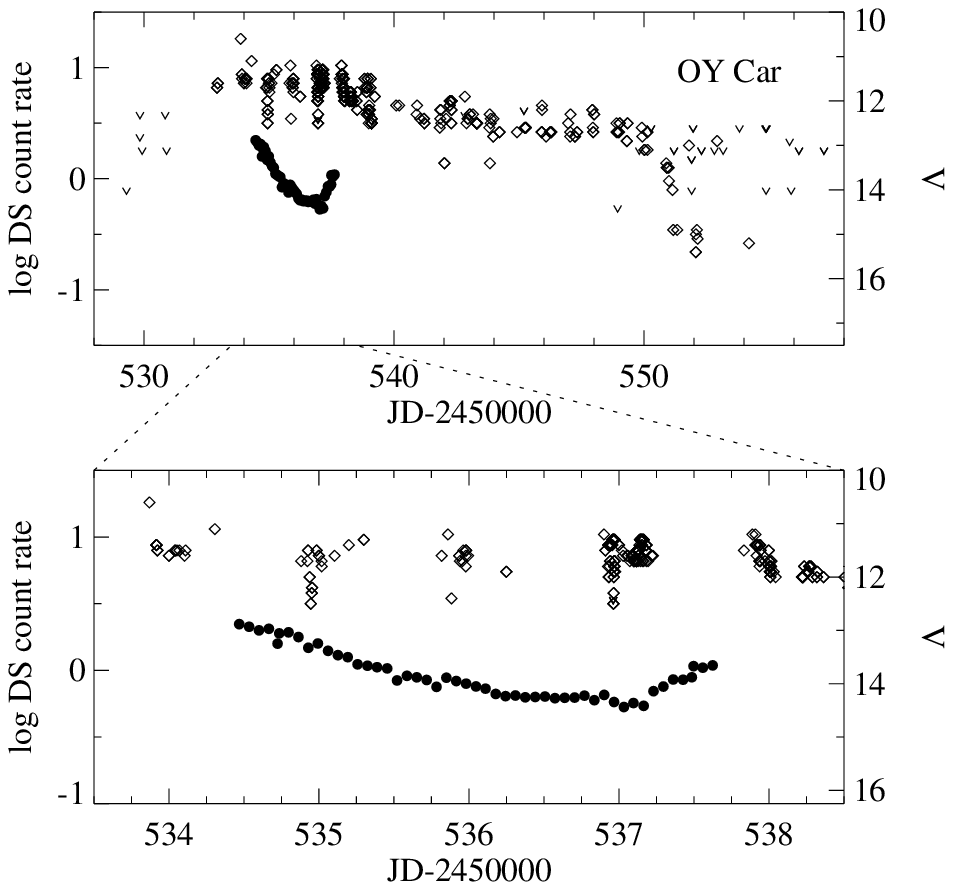}{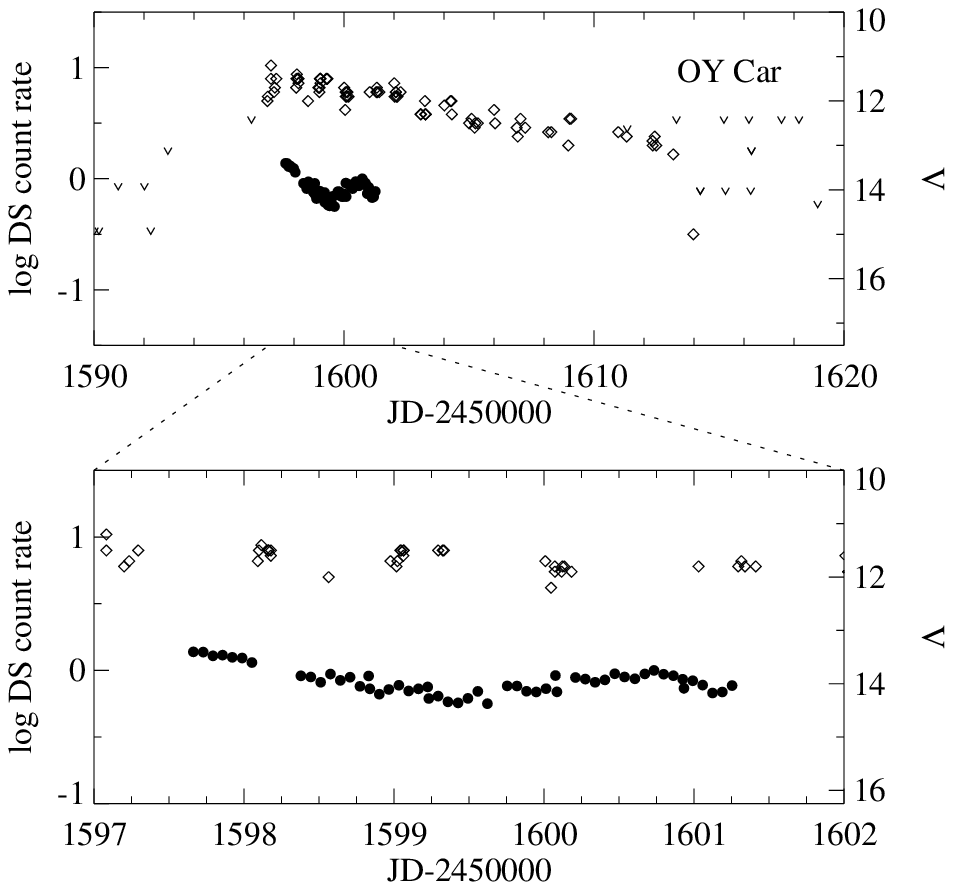}
\caption{Superoutbursts of OY Car.}
\end{figure}

\section{Discussion}

The anomalous outbursts of SS~Cyg (Fig.~1) manifest the gradual increase
of the optical and EUV light curves expected for inside-out outbursts.
The optical and EUV flux rises during the beginning of these outbursts as
the heating wave sweeps outward through the disk, causing more and more
material to flow through the disk and boundary layer onto the white dwarf.
In contrast, the normal outbursts of SS~Cyg (Fig.~2), U~Gem (Fig.~3), and
VW~Hyi (Fig.~4) manifest the fast increase of, and the delay between,
the optical and EUV light curves expected for outside-in outbursts. As
measured from the initial rise of the optical light curves, the delay of
the rise of the EUV light curves is $\approx 1.5$, 1.25, and 0.75 days for
SS~Cyg, U~Gem, and VW~Hyi, respectively. In VW~Hyi, the FUV light curve
rises $\approx 0.5$ days after the optical light curve and $\approx 0.25$
days before the EUV light curve, but falls as slowly as the optical light
curve, consistent with the expectation that the accretion disk and not the
boundary layer is the source of the FUV flux. We were not able to observe
the rise of the EUV light curves of the superoutbursts of VW~Hyi (Fig.~5)
or OY~Car (Fig.~6), but during both observations of OY~Car the EUV light
curve was observed to fall and then rise while the optical light curve
was declining only slowly. This behavior is evidence for a decrease and
subsequent increase of the mass-accretion rate onto the white dwarf, as
might be expected if a normal outburst is rejuvenated by an increase in
the mass-accretion rate driven by the tidal instability expected for
such high mass-ratio binaries.

The observed optical-EUV delays of the normal outbursts of SS~Cyg, U~Gem,
and VW~Hyi provide the most direct measurement of the velocity of the
heating wave which transforms the disk from quiescence to outburst.
Assuming the system parameters shown in Table~1, that the radius of the
disk is $R_{\rm disk}\approx 0.7\times R_{\rm L1}$, and that the disk
instability starts at the outer edge of the disk, the velocity of the
heating wave $v\approx R_{\rm disk}/{\rm delay}\approx 3~\rm km~s^{-1}$.
This result is consistent with $v=\alpha c_{\rm s}$ if the viscosity
parameter $\alpha\approx 0.2$ and the sound speed $c_{\rm s} =10\,
(T/10^4~{\rm K})^{1/2}\approx 15~\rm km~s^{-1}$.

\begin{table}
\vskip -0.2cm
\caption{System Parameters$^a$ and Velocity of the Heating Wave}
\begin{tabular}{llll}
\tableline
Parameter&	SS Cyg&	U Gem&	VW Hyi\\
\tableline
\hbox to 2.5in{$P_{\rm orb}$ (days)\leaders\hbox to 5pt{\hss.\hss}\hfil}&
	 0.2751&	0.1769&	0.0743\\
\hbox to 2.5in{$M_1$ ($M_{\sun}$)\leaders\hbox to 5pt{\hss.\hss}\hfil}&
	1.19&	  1.26&	  0.63\\
\hbox to 2.5in{$M_2$ ($M_{\sun}$)\leaders\hbox to 5pt{\hss.\hss}\hfil}&
0.70&	  0.57&	  0.11\\
\hbox to 2.5in{$q = M_1/M_2$\leaders\hbox to 5pt{\hss.\hss}\hfil}&
	1.69&	  2.17&	     6\\
\hbox to 2.5in{$R_{\rm L1}/a$\leaders\hbox to 5pt{\hss.\hss}\hfil}&
0.43&	  0.45&	  0.54\\
\hbox to 2.5in{$a = [(P_{\rm orb}/2\pi )2G(M_1+M_2)]^{1/3}$
(cm)\leaders\hbox to 5pt{\hss.\hss}\hfil}&
     $1.5\times 10^{11}$&	$1.1\times 10^{11}$&	$4.7\times 10^{10}$\\
\hbox to 2.5in{$R_{\rm disk}\approx 0.7\times R_{\rm L1}$ (cm)\leaders\hbox
to 5pt{\hss.\hss}\hfil}&
     $4.6\times 10^{10}$&	$3.5\times 10^{10}$&	$1.8\times 10^{10}$\\
\hbox to 2.5in{delay (days)\leaders\hbox to 5pt{\hss.\hss}\hfil}&
	1.5&	1.25&	0.75\\
\hbox to 2.5in{$v\approx R_{\rm disk}$/delay ($\rm km~s^{-1}$)\leaders\hbox
to 5pt{\hss.\hss}\hfil}&	3.5&	3.3&	2.7\\
\tableline
\end{tabular}
\phantom{a}$^a$Ritter, H., \& Kolb, U. 1998, \aap , 129, 83
\end{table}

\acknowledgements

We thank the many members of the AAVSO and VSS/RASNZ for providing the
triggers for our satellite observations and for the data used to construct
the optical light curves. We are grateful to J.~Cannizzo for years of
discussions about the dwarf nova instability theory and to J.~Holberg
and J.~Collins for acquiring and reducing the {\it Voyager\/} data shown
in Fig.~4. J.A.M.~acknowledges NASA grants NAG5-3144, -0727, -7357, -7358,
and 06545.03A. C.~W.~M.'s contribution to this work was performed under
the auspices of the U.S.\ Department of Energy by University of California
Lawrence Livermore National Laboratory under contract No. W-7405-Eng-48.

\end{document}